\documentstyle[12pt]{article}

\def\be{\begin{equation}}
\def\ee{\end{equation}}
\def\bea{\begin{eqnarray}}
\def\eea{\end{eqnarray}}
\def\ben{\begin{enumerate}}
\def\een{\end{enumerate}}

\def\nnu{\nonumber}
\def\ll{\label}

\begin{document}
\newcommand{\half}{{\textstyle\frac{1}{2}}}
\newcommand{\eqn}[1]{(\ref{#1})}
\newcommand{\npb}[3]{ {\bf Nucl. Phys. B}{#1} ({#2}) {#3}}
\newcommand{\pr}[3]{ {\bf Phys. Rep. }{#1} ({#2}) {#3}}
\newcommand{\prl}[3]{ {\bf Phys. Rev. Lett. }{#1} ({#2}) {#3}}
\newcommand{\plb}[3]{ {\bf Phys. Lett. B}{#1} ({#2}) {#3}}
\newcommand{\prd}[3]{ {\bf Phys. Rev. D}{#1} ({#2}) {#3}}
\newcommand{\hepth}[1]{ [{\bf hep-th}/{#1}]}
\newcommand{\grqc}[1]{ [{\bf gr-qc}/{#1}]}

\def\a{\alpha}
\def\b{\beta}
\def\g{\gamma}\def\G{\Gamma}
\def\d{\delta}\def\D{\Delta}
\def\ep{\epsilon}
\def\et{\eta}
\def\z{\zeta}
\def\t{\theta}\def\T{\Theta}
\def\l{\lambda}\def\L{\Lambda}
\def\m{\mu}
\def\f{\phi}\def\F{\Phi}
\def\n{\nu}
\def\p{\psi}\def\P{\Psi}
\def\r{\rho}
\def\s{\sigma}\def\S{\Sigma}
\def\ta{\tau}
\def\x{\chi}
\def\o{\omega}\def\O{\Omega}
\def\k{\kappa}
\def\pa {\partial}
\def\ov{\over}
\def\br{\nonumber\\}
%%%%%%%%%%%%%%%%%%%%%%%%%%%%%%
\begin{flushright}
DFPD/98/TH/39\\ hep-th/9808181
\end{flushright}
\bigskip\bigskip
\begin{center}
{\large\bf Intersecting Branes and Anti-de Sitter Spacetimes
in $SU(2)\times SU(2) 
$ Gauged Supergravity}
\vskip .9 cm
{\sc Harvendra Singh}
\footnote{e-mail: hsingh@pd.infn.it} 
 \vskip 0.05cm
INFN Sezione di Padova, Dipartimento di Fisica `Galileo Galilei' \\
Universita di Padova, Via F. Marzolo 8, 35131 Padova, Italia
\end{center}
\bigskip
\centerline{\bf ABSTRACT}
\bigskip

\begin{quote}
In this note we extend our work in a previous paper hep-th/9801038. We show
here that various intersecting brane-like configurations can be found in the 
vacuum of $D=4, N=4$ supergravity with gauged R-symmetry group $SU(2)\times 
SU(2)$. These include intersections of domain-walls, strings  and point-like
objects. Some of these 
intersecting configurations preserve $1/2$ and $1/4$ of supersymmetry. We 
observe that the previously obtained $AdS_3\times R^1$ pure axionic vacuum or 
`axio-vac' is an intersection of domain-wall with  extended string with $1/2$
supersymmetries. Also
the solutions known as `electro-vac' with geometry $AdS_2\times R^2$ can be
simply interpreted as the intersection of domain-wall with point-like object.

\end{quote}

\newpage
\section{Introduction}

In this work we will study  $N=4$ gauged  $SU(2)\times SU(2)$ 
supergravity in four dimensions \cite{fs}. 
It has been recently established that this 
gauged supergravity can be embedded into $N=1$ supergravity in 
ten dimensions as an $S^3\times S^3$ compactification \cite{cv1}. 
A previous Kaluza-Klein (KK)
interpretation for $SU(2)\times SU(2)$ gauged 
supergravity was given in \cite{abs} where this model was identified as part of 
the effective $D=4$ theory for the heterotic string in an
$S^3 \times S^3$ vacuum.   
These two Kaluza-Klein interpretations are similar 
 upto consistent truncations. For a review  on gauged supergravities and 
their KK-interpretations one can see the reports \cite{bf,dnp}.
Our present work is motivated in part by the recent studies 
on AdS-supergravities \cite{mal}
in string theory. It is an observation, at least in string inspired
low energy supergravities,
that the Freund-Rubin type vacua $AdS_{p+2}\times S^m$ \cite{fr}
 do arise in the 
near horizon limit of the spacetimes around  fundamental
$p$-branes, like M2, M5 and D3-branes. 
In addition various intersections of M-$p$-branes and D-$p$-branes 
do also give rise to spacetimes which have 
near horizon geometries as $AdS_{p+2}\times S^m\times {\cal M}^n$ 
\cite{bps}. Our aim is here to realise this observation in 
the context of above gauged
supergravity in four dimensions. There are known 
anti-de Sitter vacua like $AdS_2\times R^2$, the
`electro-vac' solution \cite{fg}, and the pure axionic background or 
`axio-vac' \footnote{This terminology 
was adopted in \cite{ct} and we shall be using it throughout this work.} 
 with the geometry $AdS_3\times R^1$ 
\cite{hs2} in four dimensional gauged supergravity.   
The idea is that the connection between $AdS$ vacua and brane geometries
must also exist in the case of gauged supergravity backgrounds. This connection
might help us to extend the $AdS/CFT$-conjecture \cite{mal} to the gauged 
supergravity sector. 
We will show explicitly that the axio-vac and electro-vac are nothing but the 
intersections of 
domain-walls with strings and point-like objects in four dimensions. 

 We start with the $D=4,~N=4$ gauged $SU(2)_A\times SU(2)_B$ supergravity 
which in the
gravity multiplet contains the graviton $E_\m^m$, four Majorana spin-$3/2$
gravitinos $\Psi^I_\m$, three non-abelian vector fields $A_\m^a$ belonging to
$SU(2)_A$, three non-abelian axial-vector fields
$B_\m^a$ belonging to $SU(2)_B$, 
four Majorana spin-$1/2$ fields $\chi^I$, the dilaton field $\phi$,
and an axion $\eta$ \cite{fs}. 
Here we consider the truncated version 
where half of the gauge fields $B_\m^a$ are vanishing. 
We will also set all the spinor background fields to zero. 
Under these specifications the  bosonic action \cite{fs} becomes
\bea
&& S= \int d^4x \sqrt{-g}~\bigg[~- R + {1\ov 2} \pa_\m \f\pa^\m\f 
+ {1\ov 2}e^{2\f} ~\pa_\m \eta\pa^\m \eta  
- {1\ov 2.2!} e^{-\f}F_{\m\n}^a F^{a\m\n}\nnu\\  
&&\hskip1cm - {1\ov 2.2!} ~ \eta~ F_{\m\n}^a \tilde F^{a\m\n}  
+{\L^2 \ov 2} e^\f \bigg]
\label{act}
\eea
where $\L^2=e_A^2+e_B^2$, $e_A$ and $e_B$ being the gauge couplings of 
respective $SU(2)$ groups and the field strength $F_{\m\n}^a$ is defined as
\be
F_{\m\n}^a=\pa_\m A_\n^a -\pa_\n A_\m^a +e_A\epsilon_{a b c} A_\m^bA_\n^c
,~~~ \tilde F_{\m\n}^a={1\ov 2} \ep_{\m\n}^{~~\s\r} F_{\s\r}^a.
\ee
The corresponding supersymmetry transformations for the vanishing fermionic
background are \cite{fs},
\bea
&&\delta\bar\chi^I={i\ov 2\sqrt{2}} \bar\ep^I\left(\pa_\m\f 
+i \g_5 e^\f\pa_\m\eta\right)\g^\m -
{1\ov 4} e^{-{\f\ov 2}}
\bar\ep^I\a^a F_{\m\n}^a \G^{\m\n} + {1\ov 4} e^{\f\ov 2}\bar\ep^I~
(e_A+i\g_5e_B),\nnu\\  
&&\delta\bar\Psi_\m^I=\bar\ep^I\left(\overleftarrow\pa_\m -{1\ov 2} 
\omega_{\m,m n}\G^{m n}
+{1\ov 2} e_A\a^a A_\m^a\right) -{i\ov 4} e^\f \bar\ep\g_5\pa_\m\eta
 -{i\ov  4\sqrt{2}} e^{-{\f\ov 2}}\bar\ep^I
\a^a F_{\n\r}^a \gamma_\m \G^{\n\r} \br &&
\hskip1cm +{i\ov 4\sqrt{2}} e^{\f\ov 2} 
\bar\ep^I (e_A+i\g_5 e_B)\g_\m,
\label{2}
\eea
where $\ep^I$ are four spacetime dependent Majorana spinors. Since fermionic
backgrounds are absent therefore  the bosonic
fields do not vary under  supersymmetry. 
Our convention for the metric is with mostly minus signs $(+---)$. 
$\{\g_\m,\g_\n\}=2~ g_{\m \n},~\m=0,1,2,3$, and $\g_5^2=1$.  
$\o_\m,^{m n}$ are the spin connections, 
and $\G_{m n}={1\ov4}
[\G_m,\G_n],$ where $m,~n$ are tangent space indices. 
$\a^a$ are three $4\times 4$ matrices belonging to $SU(2)_A$ 
which along with rest three
generators of $SU(2)_B$ generate the $({1\ov2},{1\ov2})$ 
representation of the
gauged model \cite{fs}.

As can be seen from (1) that dilaton potential has nonvanishing contribution 
from the cosmological constant term. It suggests that one can obtain a 
domain-wall (membrane) configuration in the vacuum of this theory. This 
background, independently reported in \cite{cow} and \cite{hs2}, is given by 
\bea
&&ds^2= U(y)\left( dt^2 - dx_1^2 - dx_2^2\right) - U(y)^{-1} dy^2,\nnu\\
&& \f= -ln~ U, \hskip1cm U=  m |y-y_0|,\nnu\\ 
&& A_\m^a=0,\hskip1cm \eta= 0,
\ll{3}
\eea
 $m^2= \L^2/2$. 
For this background, $e^\f$, 
analog of string coupling, vanishes at asymptotic infinity ($y\to\pm\infty$)
and so also the curvature scalar.
But both are divergent at $y=y_0$. 
Singularity at $y=y_0$ 
is the position of
the domain wall. 
The isometry group of this background is
$P_3(1,2)\times {\cal Z}_2$, where $P_3$ is three dimensional
P\"oncare group and ${\cal Z}_2$ is the reflection symmetry of the dilaton
potential around $y=y_0$. 
Domain-wall solution in  \eqn{3} 
preserves half supersymmetries with the supersymmetry parameter
for $e_B=0,~\L=e_A$ given by  
\bea
&&~\bar\ep^I~ \G_3 = i~ \bar\ep^I, \hskip1cm ~~ 
\bar\ep^I = U^{1\ov 4}~ \bar\ep_0^I
\ll{4}
\eea
$\ep_0^I$ is a constant spinor. 

Apart from domain-walls we also obtained in \cite{hs2} 
axio-vac solution with geometry $AdS_3\times R^1$ and a `dilaton-axion' 
background both of them preserve $1/4$ supersymmetries. Below we will see that
these two backgrounds are two different intersections of 
domain-walls with string.  
\section{ Orthogonally intersecting double domain-walls} 

As we saw in the previous section that
domain-wall is the most natural (fundamental) object in gauged 
supergravity which preserves $1/2$ supersymmetries, 
one can also extend the 
programme and obtain various intersections of domain walls which will
preserve less
than half supersymmetries.  
Correspondingly we obtain a background configuration representing an orthogonal
intersection of two domain walls given by
\bea 
&&ds^2=U_1 U_2 (dt^2-dx^2) - U_2 U_1^{-1} dy_1^2 -U_1U_2^{-1}dy_2^2\br
&&\f=-\log U_1U_2,
\ll{5}\eea
while other bosonic fields are set to zero. Here
$U_1=m_1|y_1|$ and $U_2=m_2|y_2|$ are the two harmonic functions along 
the transverse
directions $y_1$ and $y_2$ respectively, $m_1^2 +m_2^2=\L^2/2$. Note that
$t-x$ plain represents the common intersection plain of two walls while $y_1$
and $y_2$ serve as their mutual transverse coordinates. The background 
configuration in \eqn{5} preserves $1/4$ of the supersymmetries.

Having obtained orthogonally intersecting domain-walls it is quite 
straightforward to embed an extended string at the common 
intersection line of the two domain 
walls. In doing this we do not break any more supersymmetries and get a
new background
\bea 
&&ds^2=U_1 U_2 (dt^2-dx^2) - F U_2 U_1^{-1} dy_1^2 -F U_1U_2^{-1}dy_2^2\br
&&\f=-\log F U_1U_2, ~~~d\eta= ^*e^{-2\f} H_{(3)}\br
&&H_{(3)}=dt\wedge dx\wedge dF^{-1}
\ll{6}\eea
such that the new function $F$ depends only on $y_1$ and $y_2$ and satisfies 
the constraint
\be
(\pa_1U_1^2\pa_1+\pa_2U_2^2\pa_2)F(y_1,y_2)=0.
\ll{7}\ee
The harmonic functions $U_1$ and $U_2$ are same as in \eqn{5}.
The Killing spinors for the background \eqn{6} are, 
for $m_1=e_A/\sqrt{2},~m_2=e_B/\sqrt{2}$,
$$ \bar\ep^I= U_1^{1/4}~ U_2^{1/4}~ \bar\ep^I_0,~~~\bar\ep^I(i~\G_2+1)=0,~~~
\bar\ep^I(\G_3+\g_5)=0.$$
Before dwelling further let us discuss some specific cases below which follow
from the solutions of eq.\eqn{7}.
It is clear that there are two nontrivial solutions of the equation \eqn{7}.
 For the trivial one $F={\rm constant}$ the background \eqn{6} reduces to 
 intersecting wall configuration in \eqn{5} with `no' string. 
\begin{itemize}
\item 
The first nontrivial
configuration is obtained when we solve \eqn{7} by setting $F^{-1}=U_1 U_2$. 
Then we get
\bea 
&&ds^2=U_1 U_2 (dt^2-dx^2) - U_1^{-2} dy_1^2 -U_2^{-2}dy_2^2\br
&&\f=0, ~~~d\eta= ^*e^{-2\f} H_{(3)}\br
&&H_{(3)}=dt\wedge dx\wedge dF^{-1}.
\ll{8}\eea
We note that \eqn{8} could be the background obtained 
in near horizon limit of 
intersecting
string-five-brane description in ten dimensions \cite{ct}. 
In that case analog
of parameters $m_1$ and $m_2$ would be the charges $Q_5$ and $Q_5'$ 
of two orthogonally intersecting NS5-branes. In  background 
\eqn{8} when $m_1=m_2=\L/2$ one can define new coordinates in the
upper right ($+$ve) quardrant of the $y_1-y_2$ plain,
$$ |y_1||y_2|=e^{-\L/\sqrt{2}\r},~~~|y_1|/|y_2|=e^{\L/\sqrt{2}\s}$$
then it becomes after the rescaling $t\to \L/2~ t,~~x\to \L/2~x$
\be
ds^2=e^{-\L/\sqrt{2}\r}(dt^2-dx^2)-d\r^2-d\s^2,~~~\f=0,
~~~d\eta=-\L/\sqrt{2}~d\s
\ll{9a}
\ee 
This has a geometry of $AdS_3\times R^1$ and is similar to 
the axio-vac in \cite{hs2} as recently discussed 
in \cite{ct} in the frame work of ten dimensional theory. 

\item Instead if we consider $F^{-1}=U_2$ then we obtain from \eqn{7}
\bea
&&ds^2= U_1 \left(U_2(dt^2-dx^2)-{dy_2^2\ov U_2^2}\right)-U_1^{-1}dy_1^2\br
&&\f=\log U_1, ~~~d\eta= ^*e^{-2\f} H_{(3)}\br
&&H_{(3)}=dt\wedge dx\wedge dF^{-1}.
\ll{9}\eea
Note the three spacetime line element within bracketts in \eqn{9} is an $AdS_3$
 space. Thus what we have got is nothing but the `dilaton-axion' background of
\cite{hs2}. Further, if we set $U_1=1,~e_A=0$ in \eqn{9} then it reduces to the
 `axio-vac' where only axion field 
$\eta$ is 
nontrivial in addition to gravity. 
The geometry also becomes $AdS_3\times R^1$. It was
shown explicitly in \cite{hs2} that dilaton-axion and axio-vac 
backgrounds preserve
at least one quarter of the supersymmetry. Since for the axio-vacs
the space-time 
geometry is anti-de Sitter space there will be more Killing spinors in addition
to those reported in \cite{hs2}.  We write down complete set of Killing spinors
for the axio-vac which follows from \eqn{9} along the lines discussed above
\bea
\bar\ep^I=&& U_2^{1/4} ~\bar\ep^I_+, ~~~{\rm for}~ \bar\ep^I_+(i\G_3+1)
(\G_2\g_5+1)=0\br
 =&& \bar\ep^I_-\left(U_2^{-1/4} + {m_2\ov 2}U^{1/4}\g_5 (\G_0 t+\G_1x)
\right),~~~{\rm for}~\bar\ep^I_-(i\G_3+1)(\G_2\g_5-1)=0,
\ll{9b}\eea
where $\bar\ep^I_{\pm}$ are constant spinors. 
The fact that we are in an anti-de 
Sitter spacetime there is always an enhancement (doubling) of the 
supersymmetries \cite{bf}\footnote{I am grateful to M. Tonin for the 
discussions on this subject.}. 
This doubles the amount of supersymmetry for axio-vacs to $N=2$.
 This fact will also be at work  
 when we discuss electro-vacs in the next section
 \footnote{ We note that 
Killing spinors for various $AdS \times Sphere $ spaces
have been recently constructed in \cite{spinor}.}. 
\end{itemize}

\section{Intersections of three domain-walls}
We are also able to obtain a spacetime configuraton as the vacuum of the 
supergravity theory in \eqn{act} where all spatial directions are occupied
by domain walls. This is given by
\bea
&& ds^2= U_1U_2U_3 dt^2 -U_1^{-1}U_2U_3dy_1^2 -U_1U_2^{-1}U_3dy_2^2
-U_1U_2U_3^{-1}dy_3^2\br
&& \f=-\log U_1U_2U_3, ~~~U_i=m_i|y_i|,
\ll{10}\eea
such that $\sum_{i=1}^3m_i^2=\L^2/2$ while other bosonic fields are trivial. 
This background however breaks all
supersymmetries in the theory.  It can be supersymmetric only if one of the
functions $U_i$ is switched off in that case it reduces to the double-wall
configuration.
 Let us now dress this background with
$U(1)$ charges. To keep it simple we consider  
the case when single $U(1)$ gauge field is nontrivial and is such that
it gives rise to electric fields only. This background 
is
\bea
&& ds^2= F^{-1}U_1U_2U_3 dt^2 -F\left(
U_1^{-1}U_2U_3dy_1^2 +U_1U_2^{-1}U_3dy_2^2
+U_1U_2U_3^{-1}dy_3^2\right)\br
&& \f=-\log F U_1U_2U_3, ~~~U_i=m_i|y_i|,\br
&& F_{(2)}^a=\sqrt{2}~\delta^{a 3} dt\wedge dF^{-1}
\ll{11}\eea
where $m_i$'s satisfy the constraint as in \eqn{10} and 
the new function $F(y_1,y_2,y_3)$ satisfies
\be
( \pa_1U_1^2\pa_1 + \pa_2U_2^2 \pa_2+\pa_3 U_3^2\pa_3 )F=0
. \ll{12}\ee
Given this general situation there can be many solutions of \eqn{12}
 and hence many new backgrounds. However, not all of them will be 
supersymmetric. Consider
first the special case when $e_A=0,~m_1=e_B/\sqrt{2}$ 
and the harmonic functions 
$U_2=1$,  $U_3=1$. In that case \eqn{11} reduces to a single 
domain-wall with electric $U(1)$ flux
\bea
&& ds^2= F^{-1}U_1 dt^2 -F\left(
U_1^{-1}dy_1^2 +U_1dy_2^2
+U_1dy_3^2\right)\br
&& \f=-\log F U_1, ~~~U_1=(e_B/\sqrt{2})~|y_1|,\br
&& F_{(2)}^a=\sqrt{2} \delta^{a 3} dt\wedge dF^{-1}
\ll{13a}\eea
where $F$ satisfies 
\be
\pa_1 U_1^2\pa_1~F=0.
\ll{13b}\ee
The Killing spinors in this background are
\be
\bar\ep^I= F^{-{1\ov 4}}~ U_1^{1\ov 4}~\bar\ep_0^I  
\ll{13c}\ee
where the arbitrary constant spinors satisfy the constraints 
$\bar\ep_0^I(\G_1+\g_5)=0,~~
\bar\ep_0^I(i\a^3\G_0 +1)=0$. These two conditions on Killling spinors
break supersymmetry to $1/4$
leaving behind only four independent components hence $N=1$ supersymmetry. 
However if we consider the solution of \eqn{13b} to be $F^{-1}=U_1$ we get
\bea
&& ds^2= U_1^2dt^2 -{dy_1^2\ov U_1^2} -dy_2^2 -dy_3^2\br
&&\f =0 ,~~~ 
F_{(2)}^a=\sqrt{2} ~m_1\delta^{a 3} dt\wedge d|y_1|
\ll{13}\eea
which is the well known {\it electro-vac} solution with
geometry $AdS_2\times R^2$ obtained by
Freedmann and Gibbons\cite{fg} and preserving $N=2$ supersymmetry. 
It was shown in \cite{fg} that supersymmetric
electro-vacs exist only for purely electric configuration while in
the presence of 
magnetic field the spacetime geometry becomes $AdS_2\times S^2$ and it does 
not preserve any supersymmetry. This is the reason why we started with a
purely electric configuration in \eqn{11}. 
For the electro-vac \eqn{13} we write down the (static) Killing spinors
which straightforwardly follow from \eqn{13c} after the substitution 
$F^{-1}=U_1$
\bea 
&&\bar\ep^I = U_1^{1/2}~ \bar\ep^I_+
\br &&\bar\ep^I_+ (\G_1 +\g_5)=0, ~~~\bar\ep^I_+(1-\a^3\G_2\G_3)=0.
\ll{13d}\eea
These twin conditions, however, leave only $1/4$ supersymmetries intact. 
But the electro-vac \eqn{13} must preserve $1/2$ supersymmetries \cite{fg}.
So there must be more Killing spinors. 
One can find there does exist another set of (nonstatic) Killing spinors for 
the background \eqn{13}
\bea
&&\bar\ep^I = \bar\ep^I_- \left(U_1^{-1/2} + m_1U_1^{1/2} \g_5\G_0~ t\right)
\br &&\bar\ep^I_- (\G_1 -\g_5)=0, ~~~\bar\ep^I_-(1-\a^3\G_2\G_3)=0.
\ll{13e}\eea
which gives four more independent components. Hence Killing spinors 
\eqn{13d} and \eqn{13e} together make $N=2$ supersymmetry.

Thus the electro-vac configuration \eqn{13} describes 
a domain-wall with $U(1)$ electric-flux
along the transverse $y_1$ direction. The domain-wall occupies the directions 
$t,~y_1,~y_2$. Other solutions of eq.\eqn{12} such as $F^{-1}=U_1U_2$ and
$F^{-1}=U_1U_2U_3$ will generate further new backgrounds. 

\section{Conclusion}

In this work we have found the existence of 
various intersecting domain-wall configurations
in the vacuum of the $SU(2)\times SU(2)$ gauged supergravity in four 
dimensions. 
We have   
explicitly shown that some of the previously known $AdS$ backgrounds
of this gauged supergravity are essentially
the intersecting domain-wall
configurations. These include the identification of anti-de Sitter vacua
like `axio-vac' \cite{hs2}
 with the 
intersection of the domain-wall with an extended string and that of 
`electro-vac' \cite{fg}
 with the intersection of a domain-wall with point-like objects. An issue
relating to the supersymmetries preserved by axio-vac backgrounds is also
clarified.
We speculate that these identifications of anti-de Sitter gauged supergravity
vacua with the intersecting branes might help in extending $AdS/CFT$-conjecture
\cite{mal} down to gauged supergravity, at least in four dimensions. 

	This work also suggests that similar intersecting domain-wall 
configurations can also be obtained in other gauged supergravities in 
dimensions greater than four.   

\vskip.5cm
\noindent{\bf Acknowledgement:} This work is supported by INFN fellowship. I 
would like to thank the organisers of Miramare Summer Institute'98 at SISSA 
where part
of this work was done.


\begin{thebibliography}{99}

\bibitem{fs} D.Z. Freedman and J.H. Schwarz,\npb{137}{1978}{333}. 
\bibitem{cv1} A.H. Chamseddine and M.S. Volkov, 
{\it Non-Abelian Solitons in $N=4$
Gauged Supergravity and Leading Order String Theory}, \prd{57}{1998}{6242} ,
\hepth{9711181};
 \prl{79}{1997}{3343} \hepth{9707176}.
\bibitem{abs} I. Antoniadis, C. Bachas and A. Sagnotti, \plb{235}{1990}{255}. 
\bibitem{bf} P. Breitenlohner and D. Freedman, 
Ann. of Phys. 144 (1982) 249; 
G.W. Gibbons, C.M. Hull and P. Warner, \npb{218}{1983}{173}.
\bibitem{dnp} M.J. Duff, B.E.W. Nilsson and C.N. Pope, {\it Kaluza-Klein 
Supergravity} \pr{130}{1986}{1}.
\bibitem{mal} J. Maldacena, {\it The Large N Limit of Superconformal Field 
Theories 
and Supergravity}, \hepth{9711200}.
\bibitem{fr} P.G.O. Freund and M.A. Rubin, {\it Dynamics of dimensional 
reduction}, \plb{97}{1980}{233}.
\bibitem{bps} H.J. Boonstra, B. Peeters and K. Skenderis, {Branes and anti-de 
Sitter spacetimes}, \hepth{9801076}; {\it Brane 
intersections, anti-de Sitter spacetimes and dual superconformal theories},
\hepth{9803231}. 
\bibitem{fg} D.Z. Freedman and G.W. Gibbons, \npb{233}{1984}{24}.
\bibitem{hs2} H. Singh, {\it New supersymmetric vacua for $D=4,~N=4$ gauged
supergravity}, \plb{429}{1998}{304}, \hepth{9801038}.
\bibitem{cow} P.M. Cowdall, {\it Supersymmetric Electrovacs in Gauged 
Supergravities}, preprint R/97-51, \hepth{9710214}.
\bibitem{ct} P.M. Cowdall and P.K. Townsend, {\it Gauged Supergravity Vacua 
From Intersecting Branes}, \plb{429}{1998}{281}, \hepth{9801165}.
\bibitem{spinor} H. Lu, C.N. Pope and J. Rahmfeld, {\it A construction of 
Killing Spinors on $S^n$}, \hepth{9805151}.
\end{thebibliography}
\end{document}